\documentclass[twocolumn,showpacs,preprintnumbers,
amsmath,amssymb]{revtex4}
\usepackage{graphicx}
\usepackage{multirow}

\def\beq{\begin{equation}}
\def\eeq{\end{equation}}
\def\bea{\begin{eqnarray}}
\def\eea{\end{eqnarray}}
\def\ba{\begin{array}}
\def\ea{\end{array}}
\def\bit{\begin{itemize}}
\def\eit{\end{itemize}}

\def\nn{\nonumber}

\def\Ga{\Gamma}
\def\si{\sigma}

\def\de{\delta}

\def\rd{{\rm d}}

\def\rd{{\rm d}}

\def\cA{{\cal{A}}}
\def\cL{{\cal{L}}}

\def\GB{{{\rm E}_4}}
\makeatletter
\def\underbracket{%
  \@ifnextchar [ %
    {\@underbracket}%
    {\@underbracket [\@bracketheight]}}
\def\@underbracket[#1]{%
  \@ifnextchar [ %
    {\@under@bracket[#1]}%
    {\@under@bracket[#1][0.4em]}}
\def\@under@bracket[#1][#2]#3{%\message {Underbracket: #1,#2,#3}
  \mathop {%
    \vtop {%
      \m@th \ialign {%
        ##\crcr $\hfil \displaystyle {#3}\hfil $%
       \crcr \noalign %
       {\kern 3\p@ \nointerlineskip }%
        \upbracketfill {#1}{#2}
       \crcr \noalign %
       {\kern 3\p@ }%
     }%
   }%
  }%
  \limits%
}
\def\upbracketfill#1#2{%
  $\m@th \setbox \z@ \hbox {$\braceld$}
  \edef\@bracketheight{\the\ht\z@}\bracketend{#1}{#2}
  \leaders \vrule \@height #1 \@depth \z@ \hfill
  \leaders \vrule \@height #1 \@depth \z@ \hfill%
  \bracketend{#1}{#2}$%
}
\def\bracketend#1#2{\vrule height #2 width #1\relax}
\begin{document}
\title{Conformal Anomaly and Off-Shell Extensions of Gravity}
\author{Krzysztof A. Meissner$^1$ and Hermann Nicolai$^2$}
\affiliation{$^1$Faculty of Physics, University of Warsaw\\
ul. Pasteura 5, 02-093 Warsaw, Poland\\
$^2$Max-Planck-Institut f\"ur Gravitationsphysik
(Albert-Einstein-Institut)\\
M\"uhlenberg 1, D-14476 Potsdam, Germany\\
}

\begin{abstract} 
The gauge dependence of the conformal anomaly for spin-$\frac32$ and spin-$2$ fields
in non-conformal supergravities has been a long standing puzzle. In this Letter we argue
that the `correct' gauge choice is the one that follows from requiring all terms that would
imply a violation of  the Wess-Zumino consistency condition to be absent in the counterterm, 
because otherwise the usual link between the anomaly and the one-loop divergence becomes invalid. 
Remarkably, the `good' choice of gauge is the one that confirms our previous result \cite{MN} 
that a complete cancellation of conformal anomalies in $D=4$ can only be achieved 
for  $N$-extended (Poincar\'e) supergravities with $N\geqslant 5$.  
\end{abstract}
\pacs{04.62+v, 04.65+e}

\maketitle

\vspace{0.2cm}
\noindent{\bf 1. Introduction.} 
Conformal anomalies have been a subject of investigation for a 
long time \cite{DDI,Duff,CD,CD1,FT1,FT2,DS,BGVZ,EO,T}.
While their relevance is obvious for theories that are manifestly classically conformal,
they are equally relevant for {\em non}-conformal theories, in the same way that gauge 
anomalies are relevant even if gauge invariance is spontaneously broken. While 
the anomaly coefficients are unambiguous for spin $s\leqslant 1$ fields, it has long been 
known that in the context of (non-conformal) Poincar\'e or AdS supergravities 
the anomaly coefficients for spin-$\frac32$ and spin-2 fields suffer from an 
apparent `gauge-dependence', in the sense that they depend on how the theory is 
taken off the mass shell (defined in lowest order by $R_{\mu\nu} =0$) \cite{CD1,FT1}. 
This is a puzzling feature, because the anomaly, being a physical quantity, should {\em not} 
depend on the choice of gauge, unlike the divergent counterterms that 
must be introduced to render the results finite. This is in marked contrast 
to gauge anomalies, where such a gauge dependence is not observed, a feature that 
is known to be absolutely crucial for the consistency of the Standard Model of 
particle physics. In this Letter we propose a simple prescription to resolve this puzzle
and to arrive at a unique and physical answer also for conformal anomalies.

We refer to the papers cited above for reviews and basic results on conformal anomalies.
We assume that there exists a (non-local) effective action functional $\Ga_{\rm eff}$, 
depending on the metric $g_{\mu\nu}$ and various matter fields, such that the 
conformal anomaly ${\cal A}$ can be represented as the variational derivative
\beq\label{cA}
T^{\mu}{}_{\mu} (x)  \,\equiv \, - \frac1{\sqrt{-g(x)}} \frac{\de\Ga_{\rm eff}}{\de\si(x)}
\,= \,  {\cal A}(x) \,+\, \cdots 
\eeq
where $\si(x) \equiv \sqrt{-g(x)}$ is the conformal factor. The dots on the r.h.s. stand
for non-gravitational contributions due to matter interactions, or due to explicit or spontaneous
breaking of conformal invariance (such as explicit mass terms $\propto m^2\varphi^2$)
that are not relevant to our argument. We emphasize that (\ref{cA}) remains a perfectly
valid definition of the trace of the quantized energy momentum tensor also for 
{\em non-}conformal theories.
% and is thus is valid for {\em any} theory, whether 
%conformal or not.  
As a direct consequence of (\ref{cA}) the trace, and thus the anomaly,
must satisfy the integrability ({\em alias} Wess-Zumino) consistency condition 
\beq\label{WZ}
\frac{\de\big(\sqrt{-g} {\cal A}(x)\big)}{\de\si(y)}
\,=\, \frac{\de\big(\sqrt{-g}{\cal A}(y)\big)}{\de\si(x)}
\eeq
independently of whether we are dealing with a conformal or a non-conformal theory.

As is well known (see for instance \cite{DS}) the gravitational part of the
conformal anomaly in four dimensional space-time takes the form
\beq
{\cal A}(x) =\frac{1}{180\cdot 16\pi^2 }\Big(a \, \GB(x) \,+\, c \, C^2(x)\Big)
\label{Aeq}
\eeq
where
\bea\label{CGB}
C^2&\equiv&C_{\mu\nu\rho\si}C^{\mu\nu\rho\si}=R_{\mu\nu\rho\si}
R^{\mu\nu\rho\si}-2R_{\mu\nu}R^{\mu\nu}+\frac13 R^2\nn\\
\GB &=& R_{\mu\nu\rho\si}
R^{\mu\nu\rho\si}-4R_{\mu\nu}R^{\mu\nu}+R^2
\eea
and the coefficients $a$ and $c$ depend on the type of field coupling to gravity that
induces the anomaly.  $\GB$ is the Gauss-Bonnet density, which is a total derivative. In 
writing (\ref{Aeq}) we omit a possible further term $\propto \Box R$ which can be removed by a 
local counterterm ($\propto R^2$). It is straightforward to check that the two terms displayed 
in (\ref{Aeq}) do satisfy the consistency condition (\ref{WZ}), whereas the square of the 
scalar curvature $R^2$ by itself does not. Importantly, possible extra terms indicated by dots in
(\ref{cA}), and more specifically any terms arising for non-conformal theories, must also 
satisfy (\ref{WZ}).

Existing calculations of the conformal anomaly rely on the relation between the conformal 
anomaly and the one-loop divergence. This was in particular explained in \cite{Duff} in the 
framework of dimensional regularization, where, however, spin $s\!>\!1$ fields were not 
considered. The divergence can be generally represented in the form  
\beq
\Ga_\infty=\frac{1}{16\pi^2}\int\rd^4 x\sqrt{-g} \,\cL_\infty(x)
\eeq
where \cite{FT1}
\beq
\cL_\infty  \,=\,  b_0 L^4+b_2 L^2 +b_4\ln(L^2/\mu^2)\,+\,\ldots
\eeq
with the UV cutoff $L$, and where the dots stand for finite (non-local) contributions.
The terms $b_0$ and $b_2$ are removed by appropriate counterterms, while the 
coefficient of the logarithmic divergence contains the physically relevant information.
More precisely, using the notation from \cite{FT1} which we follow throughout,
\beq
b_4=\beta_1\GB+\frac{\beta_2}{2}(C^2-\GB)+\frac{\beta_3}{3}R^2+\cdots
\eeq
where the ellipses denote further terms {\em e.g.} involving a cosmological constant, 
as well as possible matter field contributions. In the context of supergravity 
the coefficients $\beta_i$ were computed in \cite{FT1}, building on earlier work in \cite{CD1}. 
It was also pointed out there that the counterterm coefficients 
$\beta_2$ and $\beta_3$ depend on the gauge choice, while $\beta_1$ does not.

The link between $\cL_\infty$ and the anomaly ${\cal A}$ is encapsulated in the relation
\beq\label{b4A}
%\frac{\de\Gamma_{\rm eff}}{\de\si(x)} 
\cA(x) \,\propto \,  \lim_{L\rightarrow\infty} 
\left[ L\frac{d}{dL} \big( \sqrt{-g}\cL_\infty(x)\big) \right]_{\rm F.P.} 
\eeq
where ``F.P." denotes the finite part (after subtraction of quartic and quadratic infinities).
The formula (\ref{b4A}) captures the anomalous scale dependence of the theory
in presence of a (scale breaking) cutoff $L$ (and incidentally also explains the appearance 
of $\beta$-functions on the r.h.s. of (\ref{cA}) in the presence of matter interactions).
However it is clear that the relation $b_4 = \cA$ cannot hold if $\beta_3\neq 0$ because $\beta_3$
multiplies the term $R^2$ which does not satisfy (\ref{WZ}). Because $\beta_3$ is gauge dependent 
our proposal is therefore {\em to identify the `good'  gauge as the one where $\beta_3 = 0$, 
because only then do we obtain a consistent anomaly.}

Let us now examine existing results in view of the potential discrepancy for $\beta_3\neq 0$. 
For low spins $s\leqslant 1$ the anomaly coefficients have been known for a long time 
\cite{CD1,FT1,Parker,EH,BD,Vass,BvanN}, and there is no issue (and hence no gauge dependence)
here, as the relevant actions are conformal also for non-conformal supergravities. For all
of these one finds $\beta_3 =0$. However, for spins $s\geqslant\frac32$ there are two 
different possible actions. One is conformal (super-)gravity, with kinetic terms of higher order 
(cubic for spin-$\frac32$ and fourth order for spin-2), whereas the non-conformal supergravity 
actions are of first and second order, respectively, just like the actions for spin $s\leqslant 1$. 
Higher derivative actions for higher spin fields have been investigated recently in \cite{T}, 
and for them (and thus for conformal supergravities) one gets again $\beta_3 =0$, whence 
the relation (\ref{b4A}) is preserved. For those theories it is found that a complete cancellation
of conformal anomalies is possible only for the maximal $N=4$ conformal supergravity 
coupled to an $N=4$ Yang-Mills  gauge theory with gauge group $SU(2)\times U(1)$ 
or $U(1)^4$ \cite{FT2}.

For non-conformal (Poincar{\'e} or AdS) (super)gravities the two explicit calculations with 
different off-shell formulations available in the literature give different results for spins 
$s\geqslant\frac32$, including ones with non-vanishing $\beta_3$ \cite{FT1}.  In the 
Table we show the relevant coefficients together with the values for $\beta_3$
for the two different gauges, namely the Feynman gauge \cite{FT1} and 
the harmonic gauge \cite{T}, respectively. As already  mentioned, there is no 
ambiguity for spins $s\leqslant 1$, and $\beta_3=0$. By contrast, for $s\geqslant\frac32$ 
the numbers do differ. In the first three columns of Table~1 we give the coefficients 
as extracted from \cite{T}, cf. appendix of \cite{MN}; for these $\beta_3$ vanishes 
for all spins. The coefficients in the remaining columns of the Table are based 
on Table~1 of \cite{FT1},  yielding $\beta_3= -\frac13$ for spin-$\frac32$, 
and $\beta_0= \frac34$ for spin-2.
%In table 1 of \cite{FT} we see that $\beta_3$ does vanish for (non-controversial) spins $0,\ 1/2,\ 1$ but it does not vanish for spins $3/2$ ($\beta_3=-1/3$, $c_{\frac32}=-231/2$, $a_{\frac32}=229/4$) and $2$ ($\beta_3=3/4$, $c_{2}=63$, $a_2=149$). 
We thus see that only the $a$ and $c$ coefficients from the first two columns are fully 
consistent with (\ref{WZ}), whereas the ones from the fourth and fifth columns are not. 

\renewcommand{\arraystretch}{1.7}
\begin{center}
%\begin{minipage}[t]{0.5\textwidth}
\begin{tabular}{|c||c|c|c||c|c|c|}
\hline
\multirow{2}{*}%{\diagbox{spin}{}}  
&\multicolumn{3}{c||}{\cite{T}}&\multicolumn{3}{c|}{\cite{FT1}}\\ \cline{2-7}
& $c_s$ & $a_s$&$\beta_3$& $c_s$ & $a_s$&$\beta_3$\\ \hline \hline
$0$ & $\frac32$  & $-\frac12$&0  & $\frac32$  & $-\frac12$ &0 \\[3pt] \hline
$\frac12$ & $\frac92$  & $-\frac{11}{4}$&0  & $\frac92$    & $-\frac{11}{4} $ &0\\[3pt] \hline
$1$ & $18$  & $-31$&0   & $18 $   & $- 31$&0 \\[3pt] \hline
$\frac32$ & $-\frac{411}{2}$   & $\frac{589}{4}$&0    & $-\frac{231}{2}$    & $\frac{229}{4} $ &$-\frac13$\\[3pt] \hline
$2$ & $783$   & $-571$ &$0$   & $63 $   & $149$ &$\frac34$ \\[3pt] \hline
\end{tabular}
%\end{minipage}

\vspace{2mm}
Table 1. Anomaly related coefficients for two gauges. 
\end{center}

Remarkably, the cancellations exhibited in \cite{MN}, to wit,
\bea
c_2 + 5 c_{\frac32} + 10 c_1 + 11 c_{\frac12} + 10 c_0 &=& 0\;\nn\\
c_2 + 6 c_{\frac32} + 16 c_1 + 26 c_{\frac12} + 30 c_0 &=& 0 \; ,\nn\\
c_2 + 8 c_{\frac32} + 28 c_1 + 56 c_{\frac12} + 70 c_0 &=& 0 \;,
\label{cancel}
\eea
work only with the numbers from \cite{T}, which are the ones yielding a consistent anomaly.
Inspecting the differences in the coefficients for $s\geqslant\frac32$, {\em viz.}
\beq
\Delta a_{3/2} = - \Delta c_{3/2} = - 90\;, \quad \Delta a_2 = - \Delta c_2 = 720
\eeq 
(which are in accord with the fact that only the sum $(a+c)$ is gauge invariant \cite{CD,FT1})
we see that the cancellation persists for $N=8$ supergravity (as already noted in \cite{FT1}),
but fails for lower $N$. From the present point of view, however,  this is just an accident
and the cancellations might not work for yet different ways of taking the theory off shell. 
Further clarification of the ambiguities could come from a Feynman diagram computation 
of a three-point correlator of energy-momentum tensors (as opposed to the determination 
of the conformal anomaly from the one-loop divergence, as in previous work), analogous to the textbook derivation of the axial anomaly. To the best of our knowledge such a 
calculation, which could exhibit the dependence of $\beta_3$ on a continuous gauge 
parameter, is currently not available.

The phenomenon that we describe is similar in spirit to the one discovered in \cite{KM}. There is a continuous family of $\alpha'$ actions of the string gravity (gravity + dilaton + antisymmetric tensor) that can be changed by the lower order equations of motion but only one action exhibits $O(d,d)$ symmetry \cite{MV} and this action was proposed in \cite{KM} as the 'correct' one.

We consider the cancellations in (\ref{cancel}) as indicative of a hidden conformal structure 
of unknown type underlying these theories (and possibly M theory). In \cite{MN} 
we furthermore suggested a link between these cancellations and the cancellation of 
composite anomalies in extended supergravities \cite{Marcus}, as well as the so far
unexplained finiteness of $N\geqslant 5$ supergravities \cite{Kallosh}; further evidence
for such a link has been exhibited in very recent work \cite{K1,K2}. \\[2mm]

\noindent{\bf {Acknowledgments:}}  K.A.M. thanks the AEI for hospitality and support
during this work; his research was also supported by the Polish NCN grant 
DEC-2013/11/B/ST2/04046. We are most grateful to Arkady Tseytlin for correspondence
and clarifications of his earlier work. H.N. would also like to thank Gianguido Dall'Agata 
and the Padova group for an enjoyable visit and discussions related to this work.

\end{document}